\documentclass[runningheads]{llncs}
\usepackage[T1]{fontenc}
\usepackage{graphicx}
\usepackage{listings}
\usepackage{xcolor}
\usepackage{multirow}

\begin{document}
\title{Analysing semantic data storage in Distributed Ledger Technologies for Data Spaces}
\titlerunning{Analysing semantic data storage in DLT for Data Spaces}
%
\author{Juan Cano-Benito\inst{1} \and
Andrea Cimmino\inst{2} \and
Sven Hertling\inst{1} \and
Heiko Paulheim\inst{1} \and
Raúl García-Castro\inst{2}}

\authorrunning{Cano-Benito et al.}

\institute{
Data and Web Science Group, Universität Mannheim, Germany\\
\email{\{juan.cano.de.benito, sven.hertling, heiko.paulheim\}@uni-mannheim.de}
\and
Ontology Engineering Group, Universidad Politécnica de Madrid, Spain\\
\email{\{andreajesus.cimmino, r.garcia\}@upm.es}
}

\maketitle              
\begin{abstract}
Data spaces are emerging as decentralised infrastructures that enable sovereign, secure, and trustworthy data exchange among multiple participants. To achieve semantic interoperability within these environments, the use of semantic web technologies and knowledge graphs has been proposed. Although distributed ledger technologies (DLT) fit as the underlying infrastructure for data spaces, there remains a significant gap in terms of the efficient storage of semantic data on these platforms. This paper presents a systematic evaluation of semantic data storage across different types of DLT (public, private, and hybrid), using a real-world knowledge graph as an experimental basis. The study compares performance, storage efficiency, resource consumption, and the capabilities to update and query semantic data. The results show that private DLTs are the most efficient for storing and managing semantic content, while hybrid DLTs offer a balanced trade-off between public auditability and operational efficiency. This research leads to a discussion on the selection of the most appropriate DLT infrastructure based on the data sovereignty requirements of decentralised data ecosystems.


\end{abstract}

\section{Introduction}

Data spaces emerge as decentralised infrastructures with common regulatory and technical standards, where diverse actors can share, access, and use data in a secure, reliable, and trustworthy way~\cite{curry2020future,farrell2023european}. In recent years, several data space efforts have been launched with the goal of managing and sharing data in specific domains or communities, such as mobility, manufacturing, and research~\cite{curry2020future}.. One of the core principles of data spaces is data sovereignty, an exclusive self-determination of economic data goods, which differentiates data spaces from open data portals~\cite{curry2020future,farrell2023european}. 

Different data space implementations have been proposed, but there is no 
common implementation that follows the same patterns~\cite{curry2020future}. As data spaces principles are similar to Distributed Ledger Technologies (DLT) principles, some authors have proposed the use of DLT technologies as an implementation of data spaces for data exchange and sharing between organisations~\cite{meneguzzo2023integrating,kovach2024sovereign,prinz2022blockchain,ranathunga2024integrating,witharanage2024hfb}. 

In addition, a key challenge remains underexplored: although data spaces aim to support interoperability through homogeneous, semantically enriched data~\cite{curry2020future}, many of the current implementations fall short in leveraging heterogeneous and linked data at scale. To address this problem, some recent work proposes the use of semantic web languages as a standard data model to support semantic interoperability~\cite{lu2021sustainable,bader2020international,hagedorn2025decentralised} in combination with knowledge graphs~\cite{usmani2023towards,hubauer2018use,bader2020knowledge}.

When implementing data spaces using DLTs, a significant limitation arises: storing semantic data directly on the DLT can be 
very expensive in many DLT infrastructures~\cite{cano2020benchmarking}. Moreover, there is currently no comprehensive comparison between the different types of DLTs (public, private, and hybrid), in terms of their performance, cost, and suitability for storing semantic data.

To address this gap, this paper presents a systematic evaluation of semantic data storage on multiple DLT platforms. By implementing and benchmarking semantic storage strategies in different types of DLTs, the advantages and disadvantages of each DLT architecture when storing semantic data will be evaluated.

This approach enables stakeholders to identify the most suitable DLT infrastructure for semantic-based data storage in data spaces, depending on specific requirements such as cost efficiency, scalability, privacy, or decentralisation. It also contributes to the broader goal of operationalising semantic interoperability in decentralised data ecosystems.

The rest of the article is organised as follows:
Section~\ref{sec:background} provides an overview of data spaces, DLTs, and semantic technologies; Section~\ref{sec:relatedWork} reviews existing works that combine DLT and semantic web; Section~\ref{sec:methodology} details our method for benchmarking semantic data storage across different DLT platforms; Section~\ref{sec:expsce} presents the results of our experiments; Section~\ref{sec:discussion} discusses the advantages and disadvantages of each type of DLT in data space application; and finally, Section~\ref{sec:conclusion} summarises our contributions and outlines directions for future work.

\section{Background}\label{sec:background}

\subsection{Data spaces}
Data spaces are decentralised ecosystems designed to enable the sharing of sovereign, secure, and reliable data between multiple participants~\cite{solmaz2022enabling}. Unlike traditional centralised data platforms or open data portals, data spaces are governed by a set of shared legal, organisational, and technical frameworks that preserve data sovereignty~\cite{curry2020future,farrell2023european}.

They typically rely on common standards for identity management, data models, semantics, and access control; enabling interoperability between heterogeneous systems~\cite{solmaz2022enabling}. Recent European initiatives such as GAIA-X and the International Data Space Association (IDSA) have encouraged the development of sectoral data space architectures in areas such as manufacturing, mobility, healthcare, and research~\cite{braud2021road}.

A core feature of data spaces is the decoupling of data provision from data control: data remain at the source or under the governance of the data owner and are shared on demand with authorised parties. This fosters trust, legal compliance, and flexibility, especially in regulated environments~\cite{torre2022technological}.

Despite the proliferation of conceptual frameworks, the practical applications of data spaces vary widely and challenges remain in areas such as semantic interoperability, decentralised governance, data security, and data traceability~\cite{solmaz2022enabling}.

\subsection{Semantic web}

The semantic web is an extension of the current Web that aims to make data machine-interpretable through the use of formal semantics. Its core idea is to move from centralised data to a decentralised web of interconnected data, enabling automated reasoning, data integration, and intelligent services~\cite{hogan2020semantic}.

The standard data model for representing information on the semantic web is the Resource Description Framework (RDF). This data model represents the information as subject-predicate-object, known as triple. RDF enables the expression of complex relationships between entities in a structured and interoperable way, making it suitable for integrating heterogeneous data across domains and organisations~\cite{hogan2020semantic}. 

RDF provides the basis for representing knowledge as graphs, known as knowledge graphs. Knowledge graphs are increasingly being used in data spaces to support semantic interoperability, reasoning, and context-based data discovery~\cite{usmani2023towards,hubauer2018use,bader2020knowledge}. The standard query language for querying RDF is SPARQL~\cite{hogan2020semantic}.

\subsection{Distributed Ledger Technologies}

Distributed ledger technologies (DLT) are peer-to-peer systems that allow multiple participants to maintain a shared, tamper-evident ledger. DLT implementations differ in their degree of openness, trust assumptions, and governance models. Depending on the architecture, DLTs may operate without a central authority or under restricted control by a predefined set of participants. They provide varying degrees of data integrity, transparency, and traceability through consensus mechanisms that validate and replicate transactions between nodes.

Although some authors divided DLT technologies into private and public categories~\cite{collomb2016blockchain,zhang2020analysis,fan2022performance}, other authors categorised DLT into three main types: public, private and hybrid DLTs~\cite{alkhateeb2022hybrid,farahani2021convergence,shahaab2019applicability,ge2022hybrid,shrivas2019disruptive}. Each category has advantages and disadvantages in terms of performance, scalability, privacy, and decentralisation.

\begin{itemize}
    \item \textbf{Public DLT}, also known as blockchain or permissionless ledger. Blockchain allows anyone to join the network, validate transactions, and participate in consensus. These systems are optimised for transparency and decentralisation, making them suitable for trustless environments. However, their openness poses challenges to privacy and confidentiality, particularly when sensitive data is involved. The most popular blockchains are Ethereum and Bitcoin~\cite{rankhambe2019comparative}.

    Since in public DLTs anyone can participate, transactions usually have an associated cost. In Ethereum, this transaction cost is known as gas. Gas is a measurement unit for computational work in the Ethereum blockchain and every computation in Ethereum has an associated cost. The cost is specified in units of \textit{gas}~\cite{zarir2021developing}.
    \item \textbf{Private DLT}, or permissioned ledger, restricts participation to a known set of entities. Transactions are validated by preapproved nodes, allowing greater control over access, better performance, and more efficient consensus mechanisms. In private DLTs, data is managed through separate and private transactions~\cite{alkhateeb2022hybrid,farahani2021convergence,shahaab2019applicability,ge2022hybrid,shrivas2019disruptive}. One of the most famous private DLT is Hyperledger Fabric~\cite{foschini2020hyperledger}.

    Another type of private DLT is consortium DLT. In this paper, we include consortium DLTs as private DLTs. Although consortium DLT are governed by multiple organisations (as opposed to a single entity in private DLTs), they operate under the same basic principles: restricted access, known participants, and flexible governance~\cite{shrivas2019disruptive}.
    \item \textbf{Hybrid DLT} combines elements of public and private systems, using a public DLT to expose data for transparency and a private DLT for sensitive data and governance~\cite{alkhateeb2022hybrid,farahani2021convergence,shahaab2019applicability,ge2022hybrid,shrivas2019disruptive}.
\end{itemize}

DLTs have their main structure in blocks. These blocks store metadata (timestamp, creator of the block, hashes, etc) and data. The data includes transactions and smart contracts. Smart contracts are self-executing software deployed in DLT. These smart contracts allow to enforce rules, automate workflows, and facilitate interactions between users and data in a transparent and tamper-proof manner without relying on a trusted intermediary~\cite{olivieri2024general}. 

In public DLTs, smart contracts are typically implemented using specific languages designed for blockchain platforms. This limitation is due to the need for deterministic execution on all nodes in a decentralised environment and the requirement for the code to be lightweight and isolated to avoid compromising the integrity of the entire network~\cite{olivieri2024general}. The most famous smart contract language in public DLT is Solidity, created for Ethereum~\cite{crafa2019solidity}.

In contrast, private DLTs offer more flexibility in choosing programming languages. Smart contracts can be implemented using general-purpose languages such as Java, Python, or Go, allowing easier integration with existing systems and development tools~\cite{olivieri2024general}.

\section{Related Work}\label{sec:relatedWork}

In this section, the related work on the different DLT technologies, data spaces, and knowledge graphs is analysed. There are several proposals to implement data spaces with DLT~\cite{meneguzzo2023integrating,kovach2024sovereign,prinz2022blockchain,ranathunga2024integrating,witharanage2024hfb}, but these works do not perform an analysis of the storage and querying of RDF data stored in DLT.

Several approaches store only references of triple knowledge graphs on the DLT, delegating data storage to external services~\cite{le2019incorporating,chen2021openkg,wickremasinghe2024demonstrating,ruta2017semantic,naim2019knowledge,bellomarini2020blockchains,bandara2020enrichment}. These approaches do not store RDF triples directly within the DLT infrastructure.

The work of Cano-Benito et al. analyses the performance of storing RDF directly in blocks~\cite{cano2020benchmarking}. However, this approach is inefficient in terms of writing time and reading time. Sopek et al.~\cite{sopek2018graphchain} propose an proof of concept for a new blockchain based on RDF, but the focus of this work is modelling the metadata structure of blocks, without storing data, and no analysis is provided. 

Wang et al.~\cite{wang2019decentralized} propose a method to decentralise the governance of knowledge graphs storing data on blockchain using smart contracts for crowd-sourcing validation. However, in this work, the public DLT is used to store only metadata or voting outcomes, while RDF triples are stored and processed outside the blockchain without analysis of data storage.

Therefore, this paper presents an analysis of the cost of storing and querying knowledge graphs in different DLT technologies.

\section{Method}\label{sec:methodology}

The scope of this section is to define a method to evaluate the storage and querying of knowledge graphs in DLT. In order to implement our method, the following research questions are first proposed:

\begin{itemize}
    \item RQ1: What kind of DLT technology (public, hybrid, private) offers better performance for storing knowledge graphs?
    \item RQ2: What kind of DLT technology offers better performance for querying knowledge graphs?
    \item RQ3: How does updating knowledge graphs affect writing and querying times depending on the type of DLT?
\end{itemize}

Then, to implement the method, the most relevant DLT technologies have been selected. For public DLT, Ethereum as a public DLT~\cite{rankhambe2019comparative}, and Solidity for the smart contract language have been chosen~\cite{crafa2019solidity}. For private DLT, Hyperledger Fabric has been chosen~\cite{foschini2020hyperledger}. Finally, hybrid DLT is a private DLT in which the private ledgers created are certified using a public blockchain to ensure external auditability of the consistency of the history while preserving the privacy of the data~\cite{pelosi2023hybrid}. Therefore, for hybrid DLT, a combination of Hyperledger Fabric and Ethereum will be used.

As a knowledge graph, KBPedia\footnote{http://kbpedia.com} (version 2.10) has been used to store information in each of the DLT systems. To update the knowledge graph, version 2.50 of KBPedia has been used. While version 2.10 has 1,504,364 triples, version 2.50 has 1,175,147 triples. Between one version and the other, there are 137,948 new triples, 80,696 updated triples (the same subject and predicate, but different object), and 467,165 deleted triples. All RDF triples are in turtle format.

As the different DLTs have different architectures, section \ref{sec:methodologyPubDLT} presents the method to store RDF data in public DLT, section \ref{sec:methodologyPrivDLT} for private DLT and, finally, section \ref{sec:methodologyHybDLT} for the hybrid DLT.

\subsection{Method in public DLT}\label{sec:methodologyPubDLT}

\begin{figure}[h!]
\centering
\includegraphics[scale=0.85]{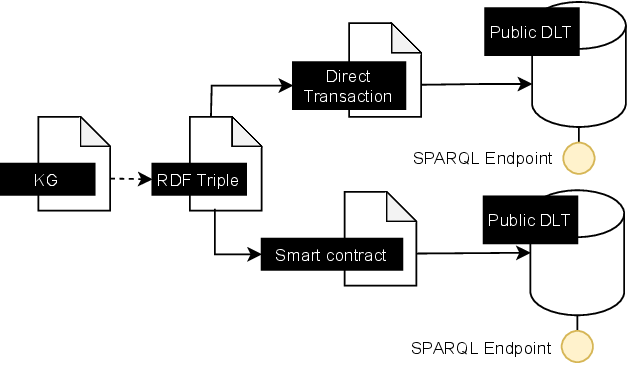}
\caption{Method in public DLT}
\label{fig:method_publicDLT}
\end{figure}

As depicted in Figure \ref{fig:method_publicDLT}, in public DLTs semantic data can be stored through two distinct approaches: (i) by embedding RDF triples directly in the data field of regular transactions and (ii) by using smart contracts.

In the direct transaction-based approach, INSERT operations are recorded by directly writing the RDF triples directly to the transaction data. DELETE operations are encoded by prefixing the triple with "DELETE", and UPDATE operations follow the format "UPDATE:OLD\_TRIPLE:NEW\_TRIPLE". These conventions enable reconstruction of the current state of the knowledge graph during querying.

In contrast, the smart contract approach provides more structured operations. Dedicated functions are implemented within the smart contract to explicitly handle insertion, deletion, and update of RDF triples. These functions modify the internal state of a contract and emit events to support traceability and synchronisation outside the chain.

Due to limitations on direct transaction size and gas consumption, RDF triples are stored individually in each transaction or smart contract. Both methods expose the stored data through an SPARQL endpoint, enabling semantic querying over the knowledge graph stored on block-chain.

\subsection{Method in private DLT}\label{sec:methodologyPrivDLT}

\begin{figure}[h!]
\centering
\includegraphics[scale=0.85]{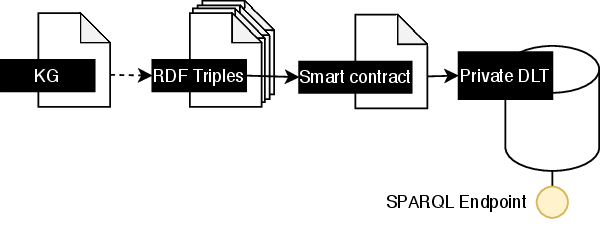}
\caption{Method in private DLT}
\label{fig:method_privateDLT}
\end{figure}

Figure \ref{fig:method_privateDLT} depicts the method to introduce semantic data in a private DLT. The RDF triples, derived from the knowledge graph, are grouped in batches (of 1000 triples each) and sent to the private ledger using smart contracts. This batch processing approach takes advantage of the flexibility of private DLT architectures, which support larger transaction sizes compared to public DLTs. This design choice reflects the technical constraints inherent in public DLTs in general, and partly explains the performance differences observed between public and private DLT implementations.

Although transaction cost is not a factor in private networks, there is a configurable maximum transaction size limit, which in Hyperledger is, by default, 49 MB\footnote{https://hyperledger-fabric.readthedocs.io/en/latest/performance.html}. This constraint will be used in this work for batch processing. Once stored, the semantic data are accessible through an SPARQL endpoint connected to the private DLT.

To manage knowledge graph updates in the private DLT, the deployed chaincode includes a method to update triples. This function receives two RDF triples: the old version and the updated version. Each old triple is first removed from the general ledger, and then the new triples are inserted. This approach allows for the actual elimination and substitution of triples within the private ledger, preserving an updated semantic view of the knowledge graph. 

\subsection{Method in hybrid DLT}\label{sec:methodologyHybDLT}

\begin{figure}[h!]
\centering
\includegraphics[scale=0.78]{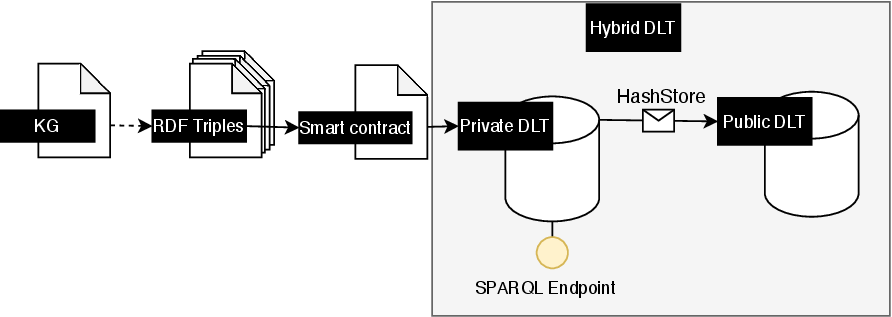}
\caption{Method in hybrid DLT}
\label{fig:method_hybridDLT}
\end{figure}

Figure \ref{fig:method_hybridDLT} depicts the method used to store RDF data in a hybrid DLT environment. In this approach, RDF triples are grouped and stored within a private DLT using smart contracts. To ensure transparency and support external auditability, a reference to this operation is recorded on a public DLT through an emitted event.

This event, named \texttt{HashStore}, is automatically triggered by the smart contract when the RDF data is registered in the private ledger. It provides a lightweight mechanism for verification and monitoring without revealing the data. The event includes the following fields:

\begin{itemize}
\item \textbf{Hash}: A cryptographic identifier generated using SHA-256 over a batch of RDF triples. This serves as a proof of integrity, enabling future verification that the data retrieved from Hyperledger Fabric remain unaltered.
\item \textbf{Submitter Address}: The Ethereum address of the entity submitting the hash. This allows tracking of the identity of the sender, which is relevant in scenarios where participant accountability is required.
\item \textbf{Timestamp}: The block timestamp where the event is recorded, providing a verifiable chronological reference.
\item \textbf{Metadata}: A descriptive string contextualising the hash, typically identifying the corresponding transaction within the private DLT.
\end{itemize}

The update mechanism in the hybrid DLT is the same as the approach used in the private DLT: first the existing triples are removed from the private ledger, and then the new versions are inserted. In addition, for each update operation, a cryptographic summary of the modified data is generated and sent to the public DLT to provide an immutable audit trail. It is important to note that the effectiveness and granularity of this public traceability may vary depending on how the smart contract is implemented and integrated in the public DLT environment.

\section{Experimental results}\label{sec:expsce}

In this section, the different scenarios are analysed. Section \ref{sec:pubdlt} analyses the storage of semantic data in the different DLT architectures, section \ref{sec:privdlt} analyses the time to read and reconstruct the stored knowledge graph, and finally sections \ref{sec:hybdlt} and \ref{sec:queryingupddlt} analyse the storage and reading of semantic data stored on the updated knowledge graph.

Public DLTs are not optimised for data storage. Each transaction consumes gas, and the immutability of the ledger prevents modification or deletion of previously stored triples. This creates challenges when updating RDF data, as it may become nontrivial to determine which triples represent the current state. In contrast, private DLTs offer more flexibility. With no strict gas or transaction size limitations, multiple RDF triples can be grouped and stored in a single transaction.

\subsection{Storing triples}\label{sec:pubdlt}

The storage of RDF triples across public, private, and hybrid DLTs reveals differences in terms of architecture, efficiency, and resource usage. Table~\ref{tab:spacePublicDLT} illustrates the usage of disk space in different DLT architectures when storing the same semantic content in RDF turtle format. Although the semantic data remain identical in all cases, the storage requirements vary significantly depending on the DLT infrastructure.

\begin{table}[]
\centering
\begin{tabular}{|l|r|r|r|r|}
\hline
\multicolumn{1}{|c|}{\textbf{\begin{tabular}[c]{@{}c@{}}RDF Triples\\ range\end{tabular}}} & \multicolumn{1}{c|}{\textbf{\begin{tabular}[c]{@{}c@{}}Public DLT\\ Direct\end{tabular}}} & \multicolumn{1}{c|}{\textbf{\begin{tabular}[c]{@{}c@{}}Public DLT\\ Smart contracts\end{tabular}}} & \multicolumn{1}{c|}{\textbf{Private DLT}} & \multicolumn{1}{c|}{\textbf{Hybrid DLT}} \\ \hline
1–100,000         & 0.17 &  3.61 & 0.06 & 0.06 \\ \hline
100,001–200,000   & 0.36 &  7.60 & 0.13 & 0.13 \\ \hline
200,001–300,000   & 0.52 & 11.70 & 0.19 & 0.19 \\ \hline
300,001–400,000   & 0.69 & 15.90 & 0.26 & 0.27 \\ \hline
400,001–500,000   & 0.86 & 20.10 & 0.32 & 0.33 \\ \hline
500,001–600,000   & 1.03 & 24.40 & 0.38 & 0.38 \\ \hline
600,001–700,000   & 1.20 & 28.70 & 0.45 & 0.45 \\ \hline
700,001–800,000   & 1.38 & 33.00 & 0.51 & 0.51 \\ \hline
800,001–900,000   & 1.55 & 37.30 & 0.58 & 0.59 \\ \hline
900,001–1,000,000 & 1.72 & 41.70 & 0.64 & 0.65 \\ \hline
1,000,001–1,100,000 & 1.90 & 46.10 & 0.71 & 0.72 \\ \hline
1,100,001–1,200,000 & 2.07 & 50.50 & 0.76 & 0.77 \\ \hline
1,200,001–1,300,000 & 2.24 & 55.00 & 0.83 & 0.84 \\ \hline
1,300,001–1,400,000 & 2.41 & 59.50 & 0.87 & 0.88 \\ \hline
1,400,001–1,500,000 & 2.58 & 64.00 & 0.93 & 0.95 \\ \hline
1,500,001–1,504,364 & 2.59 & 64.20 & 0.93 & 0.95 \\ \hline
\end{tabular}
\caption{Disk space usage in DLTs (in Gigabytes)}
\label{tab:spacePublicDLT}
\end{table}

In the direct transaction approach of the public DLT, RDF triples are serialised as strings and embedded in the data field of regular Ethereum transactions. Since this method does not alter the global state of the Ethereum Virtual Machine (EVM), it avoids the overhead of state storage and associated indexing structures. The result is a relatively compact representation that benefits from compressibility and minimal disk consumption, as the data are only included in the block payload.

In contrast, the smart contract-based approach of the public DLT introduces a larger storage footprint. Each triple must be explicitly written to the persistent storage of the contract using EVM operations. This not only increases the space required to represent each triple, but also generates event logs that record insertions, updates, or deletions. Although these logs are useful for traceability, they are retained in the public DLT log history and further contribute to the overall usage of the disk. As a result, this approach consumes more disk space despite representing the same data.

The private DLT implementation is far more efficient in terms of storage. Using platforms like Hyperledger Fabric, it allows batching of RDF triples into larger transactions, avoiding the strict constraints present in public DLT. Furthermore, because the network participants are trusted and the consensus mechanism is more lightweight, there is no need for redundant log structures or gas-efficient storage schemes. This results in substantially lower disk usage for the same data volume.

Finally, the hybrid DLT approach mirrors the storage behaviour of the private DLT for the semantic content itself. However, to ensure public verifiability and auditability, a cryptographic hash representing each batch of triples is emitted as an event in a public DLT. This introduces only a marginal increase in disk usage compared to the purely private approach, since only the hash metadata (not the triples themselves) are recorded on the public DLT.

Table~\ref{tab:timePublicDLT} compares the time required to store RDF triples in different DLT configurations. Although the semantic content is identical, the storage duration varies significantly depending on the architecture and the storage mechanism. It is important to note that in public DLTs, the time required for block propagation across the network (approximately 15 seconds per block in Ethereum~\cite{wang2021ethna}) is not included in the measurements.

\begin{table}[]
\centering
\begin{tabular}{|l|r|r|r|r|}
\hline
& \multicolumn{1}{c|}{\textbf{\begin{tabular}[c]{@{}c@{}}Public DLT\\ Direct\end{tabular}}} & \multicolumn{1}{c|}{\textbf{\begin{tabular}[c]{@{}c@{}}Public DLT\\ Smart contracts\end{tabular}}} & \multicolumn{1}{c|}{\textbf{Private DLT}} & \multicolumn{1}{c|}{\textbf{Hybrid DLT}} \\ \hline

\textbf{Maximum} & 465 & 272 & 6 & 5 \\ \hline
\textbf{Minimum} & 1  & 16 & 0.3 & 0.5 \\ \hline
\textbf{Average} & 3 & 38  & 2 & 3 \\ \hline
\textbf{Total}   & 4,449,805 & 56,808,438 & 3,151,588 & 3,873,258 \\ \hline
\end{tabular}
\caption{Time to write a transaction in a public DLTs (in ms)}
\label{tab:timePublicDLT}
\end{table}

In the public DLT using direct transactions, RDF triples are encoded as strings and placed in the data field of standard Ethereum transactions. This approach does not engage the EVM beyond basic transaction handling and avoids the overhead of contract execution and state changes. Consequently, it achieves low latency, with an average write time of just 3 milliseconds per transaction, and a total write time of approximately 4.45 million milliseconds for over 1.5 million triples.

In contrast, the public DLT with smart contracts requires executing logic inside the EVM to store each triple in a persistent state. This involves function calls, storage slot updates, and state validation across all nodes. Such operations significantly increase computational load and latency. The average write time rises to 38 milliseconds, with a total duration exceeding 56 million milliseconds for the same dataset, more than ten times the cost of the direct approach.

Private DLT offers the lowest latency, in general. Smart contracts in this environment benefit from a simplified consensus mechanism, trusted participants, and the ability to batch large sets of RDF triples into single transactions. As a result, the average write time drops to just 2 milliseconds, and the total duration for the entire dataset is roughly 3.15 million milliseconds.

The hybrid DLT configuration exhibits performance close to the private DLT. Since RDF triples are stored privately and only cryptographic hashes are recorded in the public chain, the overhead is minimal. The average write time remains low at 3 milliseconds and the total time is around 3.87 million milliseconds, only higher than the private case due to the additional operations on the public ledger.

Finally, table~\ref{tab:gasPublicDLT} summarises the gas consumption associated with the writing of RDF data to the Ethereum blockchain using different approaches. Specifically, it compares the use of direct transactions and smart contracts in public DLTs, and includes hybrid configurations. For each case, the amount of data per transaction, measured by the number of characters, is kept constant. This ensures that the observed differences in gas usage are attributable solely to the storage mechanism, rather than the payload size.

\begin{table}[]
\centering
\begin{tabular}{|l|rr|rr|r|rr|}
\hline
\multirow{2}{*}{} & \multicolumn{2}{c|}{\textbf{\begin{tabular}[c]{@{}c@{}}Public DLT\\ Direct\end{tabular}}}                                                                           & \multicolumn{2}{c|}{\textbf{\begin{tabular}[c]{@{}c@{}}Public DLT\\ Smart contracts\end{tabular}}}                                                                  & \multicolumn{1}{c|}{\textbf{\begin{tabular}[c]{@{}c@{}}Private \\ DLT\end{tabular}}} & \multicolumn{2}{c|}{\textbf{\begin{tabular}[c]{@{}c@{}}Hybrid \\ DLT\end{tabular}}}                                                                                 \\ \cline{2-8} 
                  & \multicolumn{1}{c|}{\begin{tabular}[c]{@{}c@{}}Gas \\ consumed\end{tabular}} & \multicolumn{1}{c|}{\begin{tabular}[c]{@{}c@{}}Number of \\ characters\end{tabular}} & \multicolumn{1}{c|}{\begin{tabular}[c]{@{}c@{}}Gas \\ consumed\end{tabular}} & \multicolumn{1}{c|}{\begin{tabular}[c]{@{}c@{}}Number of \\ characters\end{tabular}} & \multicolumn{1}{c|}{\begin{tabular}[c]{@{}c@{}}Not \\ applicable\end{tabular}}       & \multicolumn{1}{c|}{\begin{tabular}[c]{@{}c@{}}Gas \\ consumed\end{tabular}} & \multicolumn{1}{c|}{\begin{tabular}[c]{@{}c@{}}Number of \\ characters\end{tabular}} \\ \hline
\textbf{Max.}           & \multicolumn{1}{r|}{83,000}  & 3873 & \multicolumn{1}{r|}{2,977,492} & 3873  &  n/a & \multicolumn{1}{r|}{138,272}  &     185                                                                                 \\ \hline
\textbf{Min.}           & \multicolumn{1}{r|}{22,280}                                                  & 78                                                                                   & \multicolumn{1}{r|}{176,295}                                                 & 78                                                                                   &  n/a                                & \multicolumn{1}{r|}{138,224}                                                        &   182                                                                                   \\ \hline
\textbf{Avg.}           & \multicolumn{1}{r|}{23,1212}                                               & 130.56                                                                               & \multicolumn{1}{r|}{265,814}                                              & 130.56                                                                               &  n/a                                 & \multicolumn{1}{r|}{138,262}    &   184.26                                                                                   \\ \hline
\end{tabular}
\caption{Gas consumed in DLTs}
\label{tab:gasPublicDLT}
\end{table}

In the public DLT using direct transactions, RDF triples are stored as plain text in the data field of standard transactions. This approach does not modify the contract state or invoke complex EVM operations. Consequently, it has relatively low gas costs. The average gas consumption per transaction is approximately 23,122 units, even the maximum observed value remaining below 83,000 units.

In contrast, the smart contract-based public approach results in higher gas usage. Here, storing RDF triples requires invoking contract functions, modifying persistent storage, and often emitting events for traceability. This significantly increases the computational burden. The average gas cost exceeds 265,000 units per transaction, with maximum values reaching nearly 3 million units. Despite identical character counts per transaction, the gas cost reflects the overhead of executing and persisting data through smart contract logic.

The private DLT environment does not incur gas consumption in the same way as public Ethereum networks. Since it operates under different consensus and fee models, typically involving known and trusted nodes, gas metering is not applicable, and execution costs are not quantified using gas units. Therefore, no direct comparison is made in this context.

In hybrid DLT, RDF triples are stored within the private DLT, but a cryptographic hash representing the batch is anchored in the public DLT through an event. Although only minimal data are included in the public transaction (e.g. a hash and metadata), this operation still consumes gas. Gas usage is fixed and moderate, averaging approximately 138,261 units per event, regardless of the size of the original data batch. This reflects a consistent cost to maintain auditability without exposing semantic data. In hybrid DLT, storing information consumes more gas than the direct public DLT approach, as the information is stored in the smart contract, and this increases the cost due to the generated logs, and less than the smart contract public DLT approach as it stores non-string data.

\subsection{Querying the knowledge graph}\label{sec:privdlt}

Table~\ref{tab:timeReadPublicDLT} presents the total time required to read and reconstruct the knowledge graph of each DLT. The results show differences in retrieval performance depending on the data storage approach and the architecture of the DLT.

\begin{table}[]
\centering
\begin{tabular}{|l|c|c|l|l|}
\hline
& \multicolumn{1}{c|}{\textbf{\begin{tabular}[c]{@{}c@{}}Public DLT\\ Direct\end{tabular}}} & \multicolumn{1}{c|}{\textbf{\begin{tabular}[c]{@{}c@{}}Public DLT\\ Smart contracts\end{tabular}}} & \multicolumn{1}{c|}{\textbf{Private DLT}} & \multicolumn{1}{c|}{\textbf{Hybrid DLT}} \\ \hline
\textbf{Total time} & \multicolumn{1}{r|}{421,964}                             & \multicolumn{1}{r|}{55,191}                                       & \multicolumn{1}{r|}{41,617} & \multicolumn{1}{r|}{68,861} \\ \hline
\end{tabular}
\caption{Time to read and reconstruct the knowledge graph in DLT (in ms)}
\label{tab:timeReadPublicDLT}
\end{table}

In the case of public DLTs with direct transactions, each RDF triple must be manually retrieved by scanning the full chain history, resulting in longer processing times. In contrast, smart contract-based storage allows for faster retrieval by leveraging indexed events and internal state management, significantly reducing the time required to rebuild the knowledge graph.

Private DLTs demonstrate the best overall performance, thanks to more efficient access to key-value state storage and the absence of expensive consensus operations. Hybrid DLTs, while slightly slower than private DLTs (as reading the triples performs a verification of the triplet in the public DLT), still benefit from efficient querying on the private side, with minimal overhead introduced by data verification in the public DLT.

\subsection{Updating the knowledge graph}\label{sec:hybdlt}

Updating RDF triples varies notably across DLT types due to the underlying data model and architectural constraints. In public DLTs using direct transaction storage, updates are not applied by modifying the existing data but rather by appending new entries. These entries include flags or prefixes that indicate whether a triple has been updated or deleted, such as UPDATE: or DELETE:. This approach enables state reconstruction but increases the complexity of reading and interpreting the current version of the knowledge graph.

In contrast, smart contract-based implementations support a more structured update mechanism. Dedicated functions can be defined for the insertion, deletion, and modification of triples. These allow for precise control over the state stored in the contract and typically emit events to support auditability and synchronisation.

The storage footprint of each approach during updates is shown in Table~\ref{tab:spaceUpdateDLT}, while Table~\ref{tab:timeUpdatePublicDLT} summarises the total time required to update the knowledge graph. Public DLTs with smart contracts incur significantly higher update times as a result of the computational overhead involved in modifying the chain state. Private DLTs, on the other hand, allow rapid deletion and reinsertion of triples using batch operations, resulting in the fastest update times.

\begin{table}[]
\centering
\begin{tabular}{|l|r|r|r|r|}
\hline
\multicolumn{1}{|c|}{\textbf{\begin{tabular}[c]{@{}c@{}}RDF Triples\\ range\end{tabular}}} & \multicolumn{1}{c|}{\textbf{\begin{tabular}[c]{@{}c@{}}Public DLT\\ Direct\end{tabular}}} & \multicolumn{1}{c|}{\textbf{\begin{tabular}[c]{@{}c@{}}Public DLT\\ Smart contracts\end{tabular}}} & \multicolumn{1}{c|}{\textbf{Private DLT}} & \multicolumn{1}{c|}{\textbf{Hybrid DLT}} \\ \hline
1,500,001–1,600,000   & 2.70 &  64.60 & 0.97 & 1.01 \\ \hline
1,600,001–1,700,000   & 2.85 &  65.80 & 1.02 & 1.06 \\ \hline
1,700,001–1,800,000   & 3.02 & 66.00 & 1.05 & 1.09 \\ \hline
1,800,001–1,900,000   & 3.20 & 67.20 & 1.08 & 1.13 \\ \hline
1,900,001–2,000,000   & 3.39 & 69.80 & 1.12 & 1.18 \\ \hline
2,000,001–2,000,000   & 3.57 & 71.10 & 1.19 & 1.25 \\ \hline
2,100,001–2,190,173   & 3.68 & 71.60 & 1.25 & 1.33 \\ \hline
\end{tabular}
\caption{Disk space usage in DLTs (in Gigabytes)}
\label{tab:spaceUpdateDLT}
\end{table}

Hybrid DLTs follow the same update logic as private DLTs but include an additional step: after updating the private ledger, a cryptographic hash of the new data batch is submitted to a public DLT.

\begin{table}[]
\centering
\begin{tabular}{|c|r|r|r|r|}
\hline
\multicolumn{1}{|l|}{}           & \multicolumn{1}{c|}{\textbf{\begin{tabular}[c]{@{}c@{}}Public DLT\\ Direct\end{tabular}}} & \multicolumn{1}{c|}{\textbf{\begin{tabular}[c]{@{}c@{}}Public DLT\\ Smart contracts\end{tabular}}} & \multicolumn{1}{c|}{\textbf{\begin{tabular}[c]{@{}c@{}}Private \\ DLT\end{tabular}}} & \multicolumn{1}{c|}{\textbf{\begin{tabular}[c]{@{}c@{}}Hybrid \\ DLT\end{tabular}}} \\ \hline
\textbf{Maximum} & 132 & 285  & 3 & 5 \\ \hline
\textbf{Minimum} & 1 & 16 & 0.3 & 0.6 \\ \hline
\textbf{Average} & 3 & 29 & 2 & 3 \\ \hline
\textbf{\begin{tabular}[c]{@{}c@{}}Updating \\ (time only)\end{tabular}} & 1,850,135 & 16,992,078 & 1,156,887 & 1,901,365 \\ \hline
\textbf{\begin{tabular}[c]{@{}c@{}}Total time \\ (graph construction \\ and update)\end{tabular}} & 2,312,792 & 17,087,962 & 1,239,197 & 1,983,675 \\ \hline
\end{tabular}
\caption{Time to update a knowledge graph in DLTs (in ms)}
\label{tab:timeUpdatePublicDLT}
\end{table}

Finally, Table~\ref{tab:gasUpdatePublicDLT} reports the gas consumption associated with update operations in public and hybrid DLTs. As expected, smart contract-based updates on public DLTs require substantially more gas compared to direct transactions. The hybrid approach introduces a moderate and fixed gas overhead per update due to the hash used on the public DLT.

\begin{table}[]
\centering
\begin{tabular}{|l|rr|rr|r|rr|}
\hline
\multirow{2}{*}{} & \multicolumn{2}{c|}{\textbf{\begin{tabular}[c]{@{}c@{}}Public DLT\\ Direct\end{tabular}}}                                                                           & \multicolumn{2}{c|}{\textbf{\begin{tabular}[c]{@{}c@{}}Public DLT\\ Smart contracts\end{tabular}}}                                                                  & \multicolumn{1}{c|}{\textbf{\begin{tabular}[c]{@{}c@{}}Private \\ DLT\end{tabular}}} & \multicolumn{2}{c|}{\textbf{\begin{tabular}[c]{@{}c@{}}Hybrid \\ DLT\end{tabular}}}                                                                                 \\ \cline{2-8} 
                  & \multicolumn{1}{c|}{\begin{tabular}[c]{@{}c@{}}Gas \\ consumed\end{tabular}} & \multicolumn{1}{c|}{\begin{tabular}[c]{@{}c@{}}Number of \\ characters\end{tabular}} & \multicolumn{1}{c|}{\begin{tabular}[c]{@{}c@{}}Gas \\ consumed\end{tabular}} & \multicolumn{1}{c|}{\begin{tabular}[c]{@{}c@{}}Number of \\ characters\end{tabular}} & \multicolumn{1}{c|}{\begin{tabular}[c]{@{}c@{}}Not \\ applicable\end{tabular}}       & \multicolumn{1}{c|}{\begin{tabular}[c]{@{}c@{}}Gas \\ consumed\end{tabular}} & \multicolumn{1}{c|}{\begin{tabular}[c]{@{}c@{}}Number of \\ characters\end{tabular}} \\ \hline
\textbf{Max.} & \multicolumn{1}{r|}{68,072} & 1444 & \multicolumn{1}{r|}{1,389,137} & 1685 & n/a & \multicolumn{1}{r|}{138,372} & 185  \\ \hline
\textbf{Min.} & \multicolumn{1}{r|}{22,232} & 75 & \multicolumn{1}{r|}{25,426} & 155 & n/a & \multicolumn{1}{r|}{138,224} & 182 \\ \hline
\textbf{Avg.} & \multicolumn{1}{r|}{24,205} & 133.17 & \multicolumn{1}{r|}{83,441.48} & 133.07 &  n/a & \multicolumn{1}{r|}{138,250} & 184 \\ \hline
\end{tabular}
\caption{Gas consumed when updating knowledge graphs}
\label{tab:gasUpdatePublicDLT}
\end{table}

\subsection{Querying the updated knowledge graph}\label{sec:queryingupddlt}

After performing updates, querying the current state of the knowledge graph involves reconstructing it by retrieving only the most recent and valid RDF triples. Table~\ref{tab:timeReadUpdateDLT} presents the total time required to read and rebuild the updated knowledge graph across the different DLT configurations.

\begin{table}[]
\centering
\begin{tabular}{|l|c|c|l|l|}
\hline
& \multicolumn{1}{c|}{\textbf{\begin{tabular}[c]{@{}c@{}}Public DLT\\ Direct\end{tabular}}} & \multicolumn{1}{c|}{\textbf{\begin{tabular}[c]{@{}c@{}}Public DLT\\ Smart contracts\end{tabular}}} & \multicolumn{1}{c|}{\textbf{Private DLT}} & \multicolumn{1}{c|}{\textbf{Hybrid DLT}} \\ \hline
\textbf{Total time} & \multicolumn{1}{r|}{681,746} & \multicolumn{1}{r|}{82,605} & \multicolumn{1}{r|}{43,897} & \multicolumn{1}{r|}{51,314} \\ \hline
\end{tabular}
\caption{Time to read and reconstruct the updated knowledge graph (in ms)}
\label{tab:timeReadUpdateDLT}
\end{table}

Public DLTs that rely on smart contracts show better performance due to the structured way in which data are stored and retrieved, often leveraging indexed events or internal mappings that reduce search times. However, performance still lags behind that of private DLT due to the inherent constraints of public DLT platforms.

Private DLTs again show the best read performance, benefiting from efficient key-based access and the ability to directly discard obsolete triples during update operations. The hybrid approach introduces minimal overhead compared to the private DLT, since the querying process is still handled by the private DLT, and the public ledger is only used for integrity verification and traceability.

These results confirm that, while public DLTs ensure high transparency, they are less suitable for use cases requiring frequent updates and fast query responses over dynamic semantic data.

\section{Discussion}\label{sec:discussion}

This work provides a comparative evaluation of public, private and hybrid DLTs for storing and managing semantic data in the form of RDF triples. The results show differences between transparency, performance, scalability, and storage efficiency when designing DLT-based data spaces with knowledge graphs.

\subsection{Performance and storage efficiency}

Private DLTs outperform public and hybrid DLTs in terms of write and update performance. Due to the absence of gas consumption and the ability to batch larger transactions, a private DLT like Hyperledger Fabric enables faster processing of large volumes of RDF data with significantly lower disk usage. In contrast, public DLTs such as Ethereum incur high overhead due to transaction fees and storage costs, especially when using smart contracts for persistence. The data show that writing RDF triples using smart contracts on Ethereum consumes more gas and disk space, with longer execution times compared to using direct transactions.

\subsection{Query and update behaviour}

Querying and updating the knowledge graph also varies according to the DLT architecture. In public DLTs, the immutable nature of such DLTs makes updating RDF more difficult. Using simple transactions, updates must be encoded as new entries, which complicates data reconstruction. Smart contracts mitigate this issue by allowing explicit functions for insertion, deletion, and modification, but at the cost of increased latency and resource consumption. However, all transactions remain publicly accessible and immutable, allowing any participant to reconstruct the complete evolution of the knowledge graph and get a snapshot at any time. This transparency favours a high level of auditability and traceability.

In private and hybrid DLTs, upgrades are managed more efficiently through customised smart contract logic that supports the complete removal and replacement of RDF triples. This enables the maintenance of an up-to-date knowledge graph within the ledger. In hybrid DLTs, cryptographic hashes issued to the public ledger record the integrity of batches of data at a given point in time, but do not reveal the data itself or the exact sequence of updates. Instead, they serve as an immutable proof evidence that external auditors can use to verify that a given batch of RDF triples has not been altered since it was recorded. This approach balances privacy with external auditability, although it requires the cooperation of the private DLT administrator to disclose the data for verification. However, the effectiveness of this strategy depends on how the public smart contract is designed and integrated.

\subsection{Hybrid DLT considerations}

The hybrid DLT architecture presents a compromise between auditability and privacy. It leverages the performance of private DLTs while allowing third-party verification through the public ledger. However, the overhead of synchronising with two different infrastructures, both in terms of complexity and performance, must be taken into account.

\subsection{Suitability for data spaces}

Each DLT type presents benefits and limitations when considered for data space implementations:

\begin{itemize}
    \item \textbf{Public DLTs} offer more transparency and decentralisation, which are valuable in open, trustless environments. However, their scalability and cost-efficiency are limited when handling large volumes of semantic data.
    \item \textbf{Private DLTs} provide high performance and control, making them well suited for domain-specific or consortium-based data spaces where participants are known and data sovereignty is important.
    \item \textbf{Hybrid DLTs} are ideal for scenarios that require a balance of internal efficiency and external accountability, particularly when traceability and regulatory compliance are priorities.
\end{itemize}

\section{Conclusions}\label{sec:conclusion}

This paper presented a systematic evaluation of the storage and management of semantic data in different types of DLT, specifically public, private, and hybrid architectures. Using a real-world knowledge graph (KBPedia) and consistent RDF triple processing pipelines, a benchmark of each DLT in terms of storage, execution time, update efficiency, and query performance is presented.

The results demonstrate the following.

\begin{itemize}
    \item Private DLTs offer the best performance and storage efficiency. Their ability to process RDF data in large batches without incurring transaction costs makes them a practical choice for scalable and controlled environments.
    \item Public DLTs, while offering strong decentralisation and transparency, are significantly constrained by gas costs, storage limitations, and update complexity. Smart contracts increase traceability, but introduce high overhead in both storage and computation.
    \item Hybrid DLTs provide a valuable compromise by combining the operational advantages of private networks with the auditability of public ledgers. However, their effectiveness largely depends on the design and integration of the public-facing components.
\end{itemize}

Therefore, the optimal DLT strategy for semantic data storage depends on the specific requirements of the data space, including trust models, performance expectations, privacy restrictions, and auditability needs.

Although this study uses the most famous DLT technologies for public and private DLTs (Ethereum and Hyperledger Fabric), other implementations within each category may exhibit variations in performance or storage efficiency.

Future work will extend this analysis to other private DLT implementations, defining methods to enhance the privacy of semantic data in DLTs using W3C standards such as ODRL\footnote{https://www.w3.org/TR/odrl-model/}, applying reasoning mechanisms on knowledge graphs stored in DLT, and developing mechanisms to verify the integrity of knowledge graphs stored in DLT.

\begin{credits}
\subsubsection{\ackname} The publication of this article was funded by the University of Mannheim.
\end{credits}

%
%
%
%

\bibliographystyle{splncs04}
\bibliography{Bibliography}

\begin{thebibliography}{10}
\providecommand{\url}[1]{\texttt{#1}}
\providecommand{\urlprefix}{URL }
\providecommand{\doi}[1]{https://doi.org/#1}

\bibitem{alkhateeb2022hybrid}
Alkhateeb, A., Catal, C., Kar, G., Mishra, A.: Hybrid blockchain platforms for the internet of things (iot): A systematic literature review. Sensors  \textbf{22}(4), ~1304 (2022)

\bibitem{bader2020international}
Bader, S., Pullmann, J., Mader, C., Tramp, S., Quix, C., M{\"u}ller, A.W., Aky{\"u}rek, H., B{\"o}ckmann, M., Imbusch, B.T., Lipp, J., et~al.: The international data spaces information model--an ontology for sovereign exchange of digital content. In: International Semantic Web Conference. pp. 176--192. Springer (2020)

\bibitem{bader2020knowledge}
Bader, S.R., Grangel-Gonzalez, I., Nanjappa, P., Vidal, M.E., Maleshkova, M.: A knowledge graph for industry 4.0. In: The Semantic Web: 17th International Conference, ESWC 2020, Heraklion, Crete, Greece, May 31--June 4, 2020, Proceedings 17. pp. 465--480. Springer (2020)

\bibitem{bandara2020enrichment}
Bandara, K.Y., Thakur, S., Breslin, J.: Enrichment of blockchain transaction management with semantic triples. In: 2020 IEEE international conference on blockchain (Blockchain). pp. 188--195. IEEE (2020)

\bibitem{bellomarini2020blockchains}
Bellomarini, L., Nissl, M., Sallinger, E.: Blockchains as knowledge graphs-blockchains for knowledge graphs (vision paper). In: KR4L@ ECAI. pp. 43--51 (2020)

\bibitem{braud2021road}
Braud, A., Fromentoux, G., Radier, B., Le~Grand, O.: The road to european digital sovereignty with gaia-x and idsa. IEEE network  \textbf{35}(2), ~4--5 (2021)

\bibitem{cano2020benchmarking}
Cano-Benito, J., Cimmino, A., Garc{\'\i}a-Castro, R.: Benchmarking the efficiency of rdf-based access for blockchain environments. In: SEKE. pp. 554--559 (2020)

\bibitem{chen2021openkg}
Chen, H., Hu, N., Qi, G., Wang, H., Bi, Z., Li, J., Yang, F.: Openkg chain: A blockchain infrastructure for open knowledge graphs. Data Intelligence  \textbf{3}(2),  205--227 (2021)

\bibitem{collomb2016blockchain}
Collomb, A., Sok, K.: Blockchain/distributed ledger technology (dlt): What impact on the financial sector? Digiworld Economic Journal (103) (2016)

\bibitem{crafa2019solidity}
Crafa, S., Di~Pirro, M., Zucca, E.: Is solidity solid enough? In: International Conference on Financial Cryptography and Data Security. pp. 138--153. Springer (2019)

\bibitem{curry2020future}
Curry, E.: Future research directions for dataspaces, data ecosystems, and intelligent systems. Real-time Linked Dataspaces: Enabling Data Ecosystems for Intelligent Systems pp. 297--304 (2020)

\bibitem{fan2022performance}
Fan, C., Lin, C., Khazaei, H., Musilek, P.: Performance analysis of hyperledger besu in private blockchain. In: 2022 IEEE international conference on decentralized applications and infrastructures (DAPPS). pp. 64--73. IEEE (2022)

\bibitem{farahani2021convergence}
Farahani, B., Firouzi, F., Luecking, M.: The convergence of iot and distributed ledger technologies (dlt): Opportunities, challenges, and solutions. Journal of Network and Computer Applications  \textbf{177},  102936 (2021)

\bibitem{farrell2023european}
Farrell, E., Minghini, M., Kotsev, A., Soler~Garrido, J., Tapsall, B., Micheli, M., Posada~Sanchez, M., Signorelli, S., Tartaro, A., Bernal~Cereceda, J., et~al.: European data spaces-scientific insights into data sharing and utilisation at scale. Tech. rep., Joint Research Centre (2023)

\bibitem{foschini2020hyperledger}
Foschini, L., Gavagna, A., Martuscelli, G., Montanari, R.: Hyperledger fabric blockchain: Chaincode performance analysis. In: ICC 2020-2020 IEEE International Conference on Communications (ICC). pp.~1--6. IEEE (2020)

\bibitem{ge2022hybrid}
Ge, Z., Loghin, D., Ooi, B.C., Ruan, P., Wang, T.: Hybrid blockchain database systems: design and performance. Proceedings of the VLDB Endowment  \textbf{15}(5),  1092--1104 (2022)

\bibitem{hagedorn2025decentralised}
Hagedorn, P., Donkers, A.J., Petrova, E., K{\"o}nig, M.: Decentralised data exchange in construction data spaces using information containers. In: 2025 European Conference on Computing in Construction \& 42nd CIB W78 Conference on IT in Construction (2025)

\bibitem{hogan2020semantic}
Hogan, A.: The semantic web: two decades on. Semantic Web  \textbf{11}(1),  169--185 (2020)

\bibitem{hubauer2018use}
Hubauer, T., Lamparter, S., Haase, P., Herzig, D.M.: Use cases of the industrial knowledge graph at siemens. In: ISWC (P\&D/Industry/BlueSky). pp. 107--108 (2018)

\bibitem{kovach2024sovereign}
Kovach, A., Lanza, J., Montalvillo, L., Urbieta, A.: Sovereign iiot data exchange using dag-based dlt and international data spaces architecture. In: Proceedings of the 4th Eclipse Security, AI, Architecture and Modelling Conference on Data Space. pp. 76--85 (2024)

\bibitem{le2019incorporating}
Le-Tuan, A., Hingu, D., Hauswirth, M., Le-Phuoc, D.: Incorporating blockchain into rdf store at the lightweight edge devices. In: Semantic Systems. The Power of AI and Knowledge Graphs: 15th International Conference, SEMANTiCS 2019, Karlsruhe, Germany, September 9--12, 2019, Proceedings 15. pp. 369--375. Springer International Publishing (2019)

\bibitem{lu2021sustainable}
Lu, J., Yang, L.T., Guo, B., Li, Q., Su, H., Li, G., Tang, J.: A sustainable solution for iot semantic interoperability: Dataspaces model via distributed approaches. IEEE Internet of Things Journal  \textbf{9}(10),  7228--7242 (2021)

\bibitem{meneguzzo2023integrating}
Meneguzzo, S., Favenza, A., Gatteschi, V., Schifanella, C.: Integrating a dlt-based data marketplace with idsa for a unified energy dataspace: Towards silo-free energy data exchange within gaia-x. In: 2023 5th Conference on Blockchain Research \& Applications for Innovative Networks and Services (BRAINS). pp.~1--2. IEEE (2023)

\bibitem{naim2019knowledge}
Naim, B.A., Klas, W.: Knowledge graph-enhanced blockchains by integrating a graph-data service-layer. In: 2019 Sixth International Conference on Internet of Things: Systems, Management and Security (IOTSMS). pp. 420--427. IEEE (2019)

\bibitem{olivieri2024general}
Olivieri, L., Arceri, V., Chachar, B., Negrini, L., Tagliaferro, F., Spoto, F., Ferrara, P., Cortesi, A.: General-purpose languages for blockchain smart contracts development: A comprenhensive study. IEEE Access  (2024)

\bibitem{pelosi2023hybrid}
Pelosi, A., Felicioli, C., Canciani, A., Severino, F.: A hybrid-dlt based trustworthy ai framework. In: 2023 IEEE International Conference on Enabling Technologies: Infrastructure for Collaborative Enterprises (WETICE). pp.~1--6. IEEE (2023)

\bibitem{prinz2022blockchain}
Prinz, W., Rose, T., Urbach, N.: Blockchain technology and international data spaces. In: Designing Data Spaces: The Ecosystem Approach to Competitive Advantage, pp. 165--180. Springer (2022)

\bibitem{ranathunga2024integrating}
Ranathunga, T., Bharti, S., McGibney, A.: Integrating distributed ledger technologies into data spaces: An architectural pattern. In: International Congress on Blockchain and Applications. pp. 226--236. Springer (2024)

\bibitem{rankhambe2019comparative}
Rankhambe, B.P., Khanuja, H.K.: A comparative analysis of blockchain platforms--bitcoin and ethereum. In: 2019 5th international conference on computing, communication, control and automation (ICCUBEA). pp.~1--7. IEEE (2019)

\bibitem{ruta2017semantic}
Ruta, M., Scioscia, F., Ieva, S., Capurso, G., Di~Sciascio, E.: Semantic blockchain to improve scalability in the internet of things. Open Journal of Internet Of Things (OJIOT)  \textbf{3}(1),  46--61 (2017)

\bibitem{shahaab2019applicability}
Shahaab, A., Lidgey, B., Hewage, C., Khan, I.: Applicability and appropriateness of distributed ledgers consensus protocols in public and private sectors: A systematic review. IEEE access  \textbf{7},  43622--43636 (2019)

\bibitem{shrivas2019disruptive}
Shrivas, M.K., Yeboah, T.: The disruptive blockchain: types, platforms and applications. Texila International Journal of Academic Research  \textbf{3},  17--39 (2019)

\bibitem{solmaz2022enabling}
Solmaz, G., Cirillo, F., F{\"u}rst, J., Jacobs, T., Bauer, M., Kovacs, E., Santana, J.R., S{\'a}nchez, L.: Enabling data spaces: Existing developments and challenges. In: Proceedings of the 1st International Workshop on Data Economy. pp. 42--48 (2022)

\bibitem{sopek2018graphchain}
Sopek, M., Gradzki, P., Kosowski, W., Kuziski, D., Tr{\'o}jczak, R., Trypuz, R.: Graphchain: a distributed database with explicit semantics and chained rdf graphs. In: Companion Proceedings of the The Web Conference 2018. pp. 1171--1178 (2018)

\bibitem{torre2022technological}
Torre-Bastida, A.I., Gil, G., Mi{\~n}{\'o}n, R., D{\'\i}az-de Arcaya, J.: Technological perspective of data governance in data space ecosystems. In: Data Spaces: Design, Deployment and Future Directions, pp. 65--87. Springer International Publishing Cham (2022)

\bibitem{usmani2023towards}
Usmani, A., Khan, M.J., G.~Breslin, J., Curry, E.: Towards multimodal knowledge graphs for data spaces. In: Companion Proceedings of the ACM Web Conference 2023. pp. 1494--1499 (2023)

\bibitem{wang2019decentralized}
Wang, S., Huang, C., Li, J., Yuan, Y., Wang, F.Y.: Decentralized construction of knowledge graphs for deep recommender systems based on blockchain-powered smart contracts. IEEE Access  \textbf{7},  136951--136961 (2019)

\bibitem{wang2021ethna}
Wang, T., Zhao, C., Yang, Q., Zhang, S., Liew, S.C.: Ethna: Analyzing the underlying peer-to-peer network of ethereum blockchain. IEEE Transactions on Network Science and Engineering  \textbf{8}(3),  2131--2146 (2021)

\bibitem{wickremasinghe2024demonstrating}
Wickremasinghe, G., Frost, S., Rafferty, K., Sharma, V.: Demonstrating a hyperledger fabric-based blockchain with knowledge graphs for a supply chain ecosystem. In: 2024 IEEE International Conference on Blockchain and Cryptocurrency (ICBC). pp. 15--16. IEEE (2024)

\bibitem{witharanage2024hfb}
Witharanage, Y.R., Figueroa-Lorenzo, S., Arrizabalaga, S.: An hfb interface for the adoption of blockchain in data spaces. Computer Science \& Information Technology (CS \& IT) ISSN pp. 2231--5403 (2024)

\bibitem{zarir2021developing}
Zarir, A.A., Oliva, G.A., Jiang, Z.M., Hassan, A.E.: Developing cost-effective blockchain-powered applications: A case study of the gas usage of smart contract transactions in the ethereum blockchain platform. ACM Transactions on Software Engineering and Methodology (TOSEM)  \textbf{30}(3),  1--38 (2021)

\bibitem{zhang2020analysis}
Zhang, S., Lee, J.H.: Analysis of the main consensus protocols of blockchain. ICT express  \textbf{6}(2),  93--97 (2020)

\end{thebibliography}

\end{document}